\documentclass[12pt]{article}   
\usepackage[super,sort&compress,comma]{natbib}
\usepackage{fancyhdr}		

\usepackage[section]{placeins}   %

\usepackage{graphicx}

\makeatletter \renewcommand\@biblabel[1]{$^{#1}$} \makeatother
\newcommand{\cen}[1]{\begin{center} #1 \end{center}}

\usepackage[running,displaymath, mathlines]{lineno}

\usepackage{hyperref}
\hypersetup{ colorlinks,
    citecolor=blue,
    filecolor=blue,
    linkcolor=blue,
    urlcolor=blue
}

\usepackage{xcolor}

\definecolor{gray}{rgb}{0.6,0.6,0.6}
\definecolor{red}{rgb}{0.85,0,0}
\definecolor{green}{rgb}{0,0.85,0}
\definecolor{blue}{rgb}{0,0,0.85}
\definecolor{beige}{rgb}{0.92,0.87,0.78}
\usepackage[all]{hypcap}    

\usepackage{amsmath,amssymb,amsfonts}
\usepackage{algorithmic}
\usepackage{graphicx}
\usepackage{textcomp}
\usepackage{caption}
\usepackage{array}
\usepackage{tabularx}

\usepackage{booktabs}

\begin{document}

\begin{titlepage}
This manuscript has been accepted for publication in Medical Physics and can also be found at \url{https://aapm.onlinelibrary.wiley.com/doi/abs/10.1002/mp.15778}.
\end{titlepage}

\setlength{\baselineskip}{0.7cm}      

\cen{\sf {\Large {\bfseries A tissue-fraction estimation-based segmentation method for quantitative dopamine transporter SPECT } } \\  
\vspace*{10mm}
{ \bfseries Ziping Liu$^1$, Hae Sol Moon$^1$, Zekun Li$^1$, Richard Laforest$^2$, Joel S. Perlmutter$^{2,3}$, Scott A. Norris$^{2,3}$, Abhinav K. Jha$^{1,2}$}\\
$^1$Department of Biomedical Engineering, Washington University, St. Louis, MO 63130, United States of America. \\
$^2$Mallinckrodt Institute of Radiology, Washington University School of Medicine, St. Louis, MO 63110, United States of America. \\
$^3$Department of Neurology, Washington University School of Medicine, St. Louis, MO 63110, United States of America.
}

\pagenumbering{arabic}
\setcounter{page}{1}
\pagestyle{plain}
Address all correspondence to: Abhinav K. Jha, E-mail: a.jha@wustl.edu

\begin{abstract}

\setlength{\baselineskip}{0.7cm}      

\noindent {\bf Background:}
Quantitative measures of dopamine transporter (DaT) uptake in caudate, putamen, and globus pallidus (GP) derived from DaT-SPECT images have potential as biomarkers for measuring the severity of Parkinson disease. 
Reliable quantification of this uptake requires accurate segmentation of the considered regions. 
However, segmentation of these regions from DaT-SPECT images is challenging, a major reason being partial-volume effects (PVEs) in SPECT.
The PVEs arise from two sources, namely the limited system resolution and reconstruction of images over finite-sized voxel grids.
The limited system resolution results in blurred boundaries of the different regions. 
The finite voxel size leads to tissue-fraction effects (TFEs), i.e., voxels contain a mixture of regions. 
Thus, there is an important need for methods that can account for the PVEs, including the TFEs, and accurately segment the caudate, putamen, and GP, from DaT-SPECT images. \\
\noindent {\bf Purpose:} 
Design and objectively evaluate a fully automated tissue-fraction estimation-based segmentation method that segments the caudate, putamen, and GP from DaT-SPECT images. \\
{\bf Methods:} 
The proposed method estimates the posterior mean of the fractional volumes occupied by the caudate, putamen, and GP within each voxel of a 3-D DaT-SPECT image. 
The estimate is obtained by minimizing a cost function based on the binary cross-entropy loss between the true and estimated fractional volumes over a population of SPECT images, where the distribution of true fractional volumes is obtained from existing populations of clinical magnetic resonance images.
The method is implemented using a supervised deep-learning-based approach. \\
{\bf Results:}
Evaluations using clinically guided highly realistic simulation studies show that the proposed method accurately segmented the caudate, putamen, and GP with high mean Dice similarity coefficients $\sim$ 0.80 and significantly outperformed ($p < 0.01$) all other considered segmentation methods. 
Further, objective evaluation of the proposed method on the task of quantifying regional uptake shows that the method yielded reliable quantification with low ensemble normalized root mean square error (NRMSE) $<$ 20$\%$ for all the considered regions. 
In particular, the method yielded even lower ensemble NRMSE $\sim$ 10$\%$ for the caudate and putamen. \\
{\bf Conclusions:} 
The proposed tissue-fraction estimation-based segmentation method for DaT-SPECT images demonstrated the ability to accurately segment the caudate, putamen, and GP, and reliably quantify the uptake within these regions. 
The results motivate further evaluation of the method with physical-phantom and patient studies. 
\end{abstract}

\section{Introduction}
\label{sec: introduction}
Parkinson disease (PD) is the second-most common neurodegenerative disease. 
This disease is relentlessly progressive and is expected to affect $12$ million people worldwide by $2040$ \cite{fahn2004neurodegeneration,dorsey2018emerging,dorsey2018parkinson}. 
Dopamine transporter (DaT) single-photon emission computed tomography (SPECT) provides a mechanism to measure \textit{in vivo} pre-synaptic dopaminergic neurons that are known to degenerate in patients with PD.
Of the various DaT-based ligands \cite{emamzadeh2018parkinson}, Ioflupane I-$123$ is FDA-approved to assist with the diagnosis of parkinsonian syndromes \cite{I131_fda}. 
Conventionally, DaT-SPECT images are analyzed visually by trained readers. 
However, such qualitative analysis is subjective, and has been observed to be error-prone and suffer from intra- and inter-reader variabilities \cite{augimeri2016cada}. 
To address these issues, studies are investigating whether the quantitative analysis of DaT uptake in striatal regions of caudate and putamen can provide more useful clinical information \cite{darcourt2010eanm}. 

Further, an important unmet therapeutic need in PD is for disease-modifying therapies \cite{lang2018disease}.
While no such therapies have been identified, several are being investigated \cite{lang2018disease, aldakheel2014pathogenesis,elkouzi2019emerging}.
Biomarkers that accurately reflect disease severity are required for these studies and the need is pressing \cite{mcghee2013systematic}.
Most studies on developing such biomarkers have focused on the striatal uptake, but such measures may correlate with severity only early in the disease \cite{perlmutter2014neuroimaging, saari2017dopamine, honkanen2019no,karimi2013validation}.
Thus, there remains an important need for improved biomarkers that can measure severity throughout the range of the disease.
In this context, measurements in globus pallidus (GP) may play a key role. 
Several post-mortem studies demonstrated altered DaT levels in GP \cite{ciliax1999immunocytochemical,porritt2005dopaminergic}.
Clinical studies have observed correlations between DaT reduction in GP and resting tremor severity \cite{helmich2011pallidal}.
In addition, Whone et al.~\cite{whone2003plasticity} observed that the $^{18}$F-DOPA uptake in the GP increased in early PD, but was lost in more advanced stages of the disease. 
These observations have led to the hypothesis that plasticity of the nigropallidal pathway may help maintain a more normal pattern of pallidal output to ventral thalamus and motor cortex in early PD, and the loss of this pathway in advanced PD is a pivotal step in disease progression. 
This then suggests the variation of DaT levels in the GP at different stages of PD.
Similar findings were reported in parkinsonism secondary to occupational manganese toxicity \cite{emamzadeh2018parkinson}.
Thus, investigating the DaT uptake in the GP may provide additional insights for measuring the severity of PD.

To study the correlation between the severity of PD and quantitative DaT uptake in the caudate, putamen, and GP requires that this uptake is reliably quantified from DaT-SPECT images.
For this purpose, several quantification methods have been developed \cite{soret2003quantitative,du2005partial,du2006model}.
In all these methods, reliable quantification of uptake within the caudate, putamen, and GP requires that these regions are segmented accurately. 
However, segmentation in SPECT is challenging, a major reason being partial-volume effects (PVEs) \cite{soret2007partial}.
The PVEs arise from two sources, namely the limited system resolution and tissue-fraction effects (TFEs) \cite{rousset2007partial}.
The limited system resolution results in blurred boundaries of the different regions. 
The resolution of clinical brain SPECT systems is typically $\sim 8~\mathrm{mm}$, and the resolution in reconstructed images can be $\sim 12~\mathrm{mm}$ \cite{stam2018performance,national1996single}.
Thus, structures such as the caudate, putamen, and GP that are close to each other, are heavily affected by this limited resolution.
Further, due to the reconstruction of images over finite-sized voxel grids, voxels on the boundary of two regions likely contain more than one region.
While this is generally an issue in medical imaging, the TFEs are more prominent in SPECT due to the larger voxel size.
Segmentation is further challenged by the small sizes of the caudate, putamen, and GP.
The volume of the caudate or putamen is $\sim 10~\mathrm{cm}^3$ and the GP is only $\sim 3~\mathrm{cm}^3$.
In fact, the GP is visually almost impossible to manually delineate from SPECT images.

One approach to segment the caudate, putamen, and GP from DaT-SPECT images is to have trained readers delineate these regions manually. 
However, manual segmentation is labor-intensive, time-consuming, and suffers from intra- and inter-reader variabilities \cite{badiavas2011spect}.
To address these issues, computer-aided SPECT segmentation methods have been developed, including those based on thresholding \cite{erdi1995threshold,grimes2012accuracy}, edge detection \cite{long19912d}, region growing \cite{Slomka:1995}, and clustering \cite{Mignotte:2001,jha2017unsupervised}. 
However, results from this study show that these methods typically segmented the caudate, putamen, and GP as highly overlapped regions.
These methods could then yield erroneous measures of regional uptake and consequently, lead to limited clinical utility \cite{matesan2018123}.
Further, to the best of our knowledge, there are no validated tools to segment the GP from DaT-SPECT images. 
Thus, there is an important need for methods that can accurately segment the caudate, putamen, and GP. 

A major reason behind the limited accuracy of existing segmentation methods is their inability to account for the PVEs, and in particular, the TFEs.
In most of these methods, segmentation is defined as a voxel-wise classification problem, i.e., each image voxel is classified as belonging to a specific region. 
These classification-based segmentation methods thus are inherently limited in accounting for the TFEs.
This inherent limitation is also observed in commonly developed deep-learning (DL)-based segmentation methods \cite{leung2020physics,lin2020deep}, which are designed and trained to classify each voxel as belonging to a certain region. 
While these methods can output a probabilistic estimate of each voxel belonging to a certain region, this probability is only a measure of uncertainty in classification and thus has no relation to TFEs. 
Similarly, other probabilistic approaches such as fuzzy \cite{chen2019incorporating} and atlas-based \cite{lee2005analysis} segmentation methods can estimate the probability of each voxel belonging to a region.
However, again, this probability is unrelated to TFEs.

To address the limitation in accounting for the TFEs while performing segmentation, Liu et al.~\cite{liu2021bayesian} recently proposed an estimation-based approach in the context of segmenting the primary tumor from oncological positron emission tomography (PET) images of patients with lung cancer. 
This approach is a Bayesian method that estimates the fractional volume occupied by the tumor within each voxel of a PET image.  
While the method demonstrated accurate segmentation performance, it suffered from limitations. 
First, the method assumes that for a certain set of patients, the true fractional volumes are known, as can be derived from manual segmentations.
However, these segmentations may often be unavailable. 
For example, in DaT-SPECT segmentation, manual delineation of the GP is very difficult, if not impossible to obtain.
Next, the method was implemented in $2$-D and thus required manual supervision to ensure that the tumor was present in each input image slice.
Thus, the method was not fully automated. 

The first major contribution of this paper is to advance on the idea of estimation-based segmentation proposed by Liu et al. to develop a Bayesian tissue-fraction estimation method for fully automated and simultaneous segmentation of the caudate, putamen, and GP, from $3$-D DaT-SPECT images.
Further, the proposed method provides a mechanism to obtain the true fractional volumes of the considered regions from other imaging modalities, thus not requiring manual segmentation of the SPECT images for obtaining the ground truth.
The second major contribution is to propose a new strategy to objectively evaluate segmentation methods on the task of regional uptake quantification.
In DaT SPECT, segmentation is performed for the clinical task of quantifying the mean uptake in the various regions of the brain.
Thus, segmentation methods should be evaluated based on this quantification task. 
Towards this goal, we propose a projection-domain-quantification-based strategy to perform such objective task-based evaluation in an optimal manner.

\section{Methods}

\subsection{Problem formulation}

Consider a patient being injected with a DaT-based tracer such as $^{123}$I-Ioflupane. 
Denote the tracer distribution within the patient by $f(\mathbf{r})$, where $\mathbf{r} = (x,y,z)$ is a $3$-D vector denoting spatial coordinates. 
Consider a SPECT system, denoted by $\mathcal{H}$, that images this patient and yields projection data, denoted by an $M$-dimensional vector, $\mathbf{g}$. 
This data is then input to a reconstruction algorithm, denoted by an operator, $\mathcal{R}$, yielding a reconstructed image, denoted by a $P$-dimensional vector, $\hat{\mathbf{f}}$. 
The tracer distribution, $f(\mathbf{r})$, is assumed to lie within the Hilbert space of square-integrable functions, denoted by $\mathbb{L}_2 (\mathbb{R}^3)$. 
The projection data, $\mathbf{g}$, and reconstructed image, $\hat{\mathbf{f}}$, are assumed to lie within the Hilbert space of Euclidean vectors, denoted by $\mathbb{E}^M$ and $\mathbb{E}^P$, respectively. 
Thus, the process of obtaining the reconstructed image from the original tracer distribution can be described as a transformation: $\mathbb{L}_2 (\mathbb{R}^3) \rightarrow \mathbb{E}^M \rightarrow \mathbb{E}^P$. 
The $\hat{\mathbf{f}}$ can be given in operator notation as
\begin{linenomath*}
\begin{equation}
    \hat{\mathbf{f}} = \mathcal{R} \left\{ \mathbf{g} \right\} = \mathcal{R} \left\{ \mathcal{H} \mathbf{f} + \mathbf{n} \right\},
\label{eq: fhat}
\end{equation}
\end{linenomath*}
where $\mathbf{n}$ is an $M$-dimensional vector denoting the Poisson-distributed noise in the SPECT system.

In this problem, $K=7$ regions of interest were considered to be segmented, namely the left and right caudate, putamen, and GP, and rest of the brain. 
The rest of the brain was referred to as background. 
Denote the support of the $k^\mathrm{th}$ region by $\phi_k(\mathbf{r})$, such that
\begin{linenomath*}
\begin{align}
    \phi_k(\mathbf{r})  = 
    \begin{cases}   1, &\text{if region $k$ occupies location $\mathbf{r}$.} \\
                    0, &\text{otherwise.}
    \end{cases}
    \label{eq: phi_k(r)}
\end{align}
\end{linenomath*}
The uptake within the $k^\mathrm{th}$ region, denoted by $\lambda_k$, is then given by
\begin{linenomath*}
\begin{align}
    \lambda_k 
    = \frac{\int d^3 \mathbf{r} \ \phi_k(\mathbf{r}) f(\mathbf{r})}
    {\int d^3 \mathbf{r} \ \phi_k(\mathbf{r})},
\label{eq: lambda_k}
\end{align}
\end{linenomath*}
where the numerator and denominator calculates the total uptake and volume within the $k^{\mathrm{th}}$ region, respectively. 

From Eq.~\eqref{eq: lambda_k}, reliable quantification of $\lambda_k$ requires accurate definition of $\phi_k(\mathbf{r})$. 
This $\phi_k(\mathbf{r})$ is typically approximated by segmenting the reconstructed SPECT image, $\hat{\mathbf{f}}$. 
Conventional SPECT segmentation methods often approximate $\phi_k(\mathbf{r})$ by assigning each voxel of $\hat{\mathbf{f}}$ as belonging to a specific region $k$.
However, due to the prominence of the TFEs in SPECT, a voxel can contain more than one region.
Thus, these classification-based segmentation methods suffer from the inability to model the TFEs. 
To address this issue, we propose a tissue-fraction estimation-based segmentation method that estimates the fractional volume occupied by each considered region within each voxel of the reconstructed image. 

Denote the voxel-basis function for the SPECT image, $\hat{\mathbf{f}}$, by $\psi^{S}_p(\mathbf{r})$ such that
\begin{linenomath*}
\begin{align}
    \psi^{S}_p(\mathbf{r})  = 
    \begin{cases}   1, &\text{if $\mathbf{r}$ lies within the $p^\mathrm{th}$ voxel of $\hat{\mathbf{f}}$.} \\
                    0, &\text{otherwise.}
    \end{cases}
    \label{eq: vox fn SPECT}
\end{align}
\end{linenomath*}
Denote the volume of each voxel of $\hat{\mathbf{f}}$ by $V$. Using Eqs.~\eqref{eq: phi_k(r)} and \eqref{eq: vox fn SPECT}, the fractional volume occupied by the $k^\mathrm{th}$ region within the $p^\mathrm{th}$ voxel of $\hat{\mathbf{f}}$, denoted by $v_{k,p}^{ideal}$, is given by
\begin{linenomath*}
\begin{align}
    v_{k,p}^{ideal} = \frac{1}{V}\int d^3\mathbf{r} \ \phi_k(\mathbf{r}) \psi^S_p(\mathbf{r}). 
\label{eq: v_{k,p}}
\end{align}
\end{linenomath*}

Estimating $v_{k,p}^{ideal}$ from $\hat{\mathbf{f}}$ is an ill-posed problem due to the null spaces of $\mathcal{H}$ and $\mathcal{R}$ operators (Eq.~\ref{eq: fhat}).
One way to alleviate this ill-posedness is to take a Bayesian approach that incorporates the prior distribution of $v_{k,p}^{ideal}$. 
From Eq.~\eqref{eq: v_{k,p}}, obtaining this prior distribution requires the knowledge of $\phi_k(\mathbf{r})$ (Eq.~\ref{eq: phi_k(r)}), which is typically unavailable.
Here we recognize that the caudate, putamen, and GP are distinguishable on T$1$-weighted MR images due to the higher spatial resolution of MR systems. 
Thus, segmenting these MR images yields a discrete but high-resolution representation of $\phi_k(\mathbf{r})$.

Denote the MR image by an $N$-dimensional vector, $\hat{\mathbf{f}}^{MR}$, where $N > P$. 
Similar to Eq.~\eqref{eq: vox fn SPECT}, denote the voxel-basis function for $\hat{\mathbf{f}}^{MR}$ by $\psi^{MR}_n(\mathbf{r})$, such that
\begin{linenomath*}
\begin{align}
    \psi^{MR}_n(\mathbf{r})  = 
    \begin{cases}   1, &\text{if $\mathbf{r}$ lies within the $n^\mathrm{th}$ voxel of $\hat{\mathbf{f}}^{MR}$.} \\
                    0, &\text{otherwise.}
    \end{cases}
    \label{eq: vox fn MR}
\end{align}
\end{linenomath*}
Consider that all the $K$ regions have been segmented from the MR image. 
Then, for the $k^\mathrm{th}$ region, define an $N$-dimensional vector $\boldsymbol{\phi}_k^{MR}$, which denotes the segmentation of that region from the MR image. 
The $n^\mathrm{th}$ element of this vector is given by
\begin{linenomath*}
\begin{align}
\phi^{MR}_{k,n} = \begin{cases} 1, &\text{if the $n^\mathrm{th}$ voxel of $\hat{\mathbf{f}}^{MR}$ is assigned to region $k$.} \\
0, &\text{otherwise.}
\end{cases}
\label{eq: phi_kn^MR}
\end{align}
\end{linenomath*}
Further, consider that the MR and SPECT images are co-registered.
Using Eqs.~\eqref{eq: vox fn SPECT}, \eqref{eq: vox fn MR}, and \eqref{eq: phi_kn^MR}, the fractional volume occupied by the $k^{\mathrm{th}}$ region within the $p^{\mathrm{th}}$ voxel of the SPECT image can be calculated as follows:
\begin{linenomath*}
\begin{align}
    v_{k,p} = 
    \frac{1}{V}  \sum_{n=1}^N \phi^{MR}_{k,n} 
    \int d^3 \mathbf{r} \ \psi^{MR}_{n} (\mathbf{r}) \psi^S_p(\mathbf{r}), 
\label{eq: v_{k,p} MR}
\end{align}
\end{linenomath*}
where the integral calculates the volume that the ${n}^{\mathrm{th}}$ voxel of the MR image occupies within the $p^{\mathrm{th}}$ voxel of the SPECT image. 

In the rest of this manuscript, the MR-defined fractional volume, $v_{k,p}$, is considered as the surrogate for the ground truth, $v_{k,p}^{ideal}$ (Eq.~\ref{eq: v_{k,p}}). 
The objective is to estimate the true fractional volumes, $\{ v_{k,p}, k=1, 2, \ldots K, p=1, 2, \ldots P \}$, from the SPECT image.

\subsection{Proposed method}
\label{sec: Methods (tissue-fraction estimation technique)}
Denote the estimate of $v_{k,p}$ by $\hat{v}_{k,p}$. 
Further, denote the vector $\{v_{k,p}, p = 1,2, \ldots, P\}$ by $\mathbf{v}_k$ and the estimate of $\mathbf{v}_k$ by $\hat{\mathbf{v}}_k$. 
A cost function is defined to penalize the deviation of $\hat{\mathbf{v}}_k$ from $\mathbf{v}_k$. 
In this problem, we choose the binary cross-entropy (BCE) loss as the basis of the cost function since this loss automatically incorporates the constraint that the values of $v_{k,p}$ and $\hat{v}_{k,p}$ lie between $0$ and $1$ \cite{creswell2017denoising,liu2021bayesian}. 
The BCE loss, denoted by $l_{BCE}(v_{k,p},\hat{v}_{k,p})$, is given by
\begin{linenomath*}
\begin{equation}
    l_{BCE}(v_{k,p},\hat{v}_{k,p}) 
    =
    -v_{k,p}\mathrm{log}(\hat{v}_{k,p}) - (1-v_{k,p})\mathrm{log}(1-\hat{v}_{k,p}).
\label{eq: l_BCE}
\end{equation}
\end{linenomath*}
The cost function, denoted by $C(\mathbf{v}_k, \hat{\mathbf{v}}_k)$, is defined as the negative of aggregate BCE loss between $v_{k,p}$ and $\hat{v}_{k,p}$ over all $P$ voxels, averaged over the ensemble of true values, $\{v_{k,p}\}$, and the noise in $\mathbf{\hat{f}}$ arising from the imaging process:
\begin{linenomath*}
\begin{equation}
    C(\mathbf{v}_k, \hat{\mathbf{v}}_k) 
    = -\sum_{p=1}^P \int d^P \hat{\mathbf{f}} \int dv_{k,p} \ \mathrm{pr}(\hat{\mathbf{f}},v_{k,p}) l_{BCE}(v_{k,p},\hat{v}_{k,p}).
\label{eq: C(v_k,vhat_k)}
\end{equation}
\end{linenomath*}
Using conditional probability to expand $\mathrm{pr}(\hat{\mathbf{f}},v_{k,p})$ and replacing the expression from Eq.~\eqref{eq: l_BCE} in Eq.~\eqref{eq: C(v_k,vhat_k)} yields
\begin{linenomath*}
\begin{equation}
\begin{split}
    &C(\mathbf{v}_k, \hat{\mathbf{v}}_k) \\
    &= \sum_{p=1}^P \int d^P \hat{\mathbf{f}} \ \mathrm{pr}(\hat{\mathbf{f}}) \int dv_{k,p} \ \mathrm{pr}(v_{k,p}|\hat{\mathbf{f}}) \big[ v_{k,p}\mathrm{log}(\hat{v}_{k,p}) + (1-v_{k,p})\mathrm{log}(1-\hat{v}_{k,p}) \big] \\
    &= \sum_{p=1}^P \int d^P \hat{\mathbf{f}} \ \mathrm{pr}(\hat{\mathbf{f}}) \int dv_{k,p} \ \mathrm{pr}(v_{k,p}|\hat{\mathbf{f}}) \big[ \mathrm{log}(1-\hat{v}_{k,p}) + v_{k,p} \{ \mathrm{log}(\hat{v}_{k,p}) - \mathrm{log}(1-\hat{v}_{k,p}) \}\big].
\end{split}
\label{eq: C(v_k,vhat_k) before simp}
\end{equation}
\end{linenomath*}
Since $\int dv_{k,p} \ \mathrm{pr}(v_{k,p}|\hat{\mathbf{f}}) = 1$, Eq.~\eqref{eq: C(v_k,vhat_k) before simp} becomes 
\begin{linenomath*}
\begin{equation}
\begin{split}
    &C(\mathbf{v}_k, \hat{\mathbf{v}}_k) \\
    &= \sum_{p=1}^P \int d^P \hat{\mathbf{f}} \ \mathrm{pr}(\hat{\mathbf{f}}) \bigg[ \mathrm{log}(1-\hat{v}_{k,p}) + 
    \{ \mathrm{log}(\hat{v}_{k,p}) - \mathrm{log}(1-\hat{v}_{k,p}) \} \int dv_{k,p} \ \mathrm{pr}(v_{k,p}|\hat{\mathbf{f}}) v_{k,p} \bigg].
\end{split}
\label{eq: C(v_k,vhat_k) after simp}
\end{equation}
\end{linenomath*}
To find the value of $\hat{v}_{k,p}$ that minimizes this cost function, we differentiate Eq.~\eqref{eq: C(v_k,vhat_k) after simp} with respect to $\hat{v}_{k,p}$ and set the derivative equal to $0$. 
Since the term $\mathrm{pr}(\hat{\mathbf{f}})$ is always non-negative, Eq.~\eqref{eq: C(v_k,vhat_k) after simp} is minimized by setting 
\begin{linenomath*}
\begin{equation}
    \frac{\partial}{\partial \hat{v}_{k,p}} \Bigg[ \mathrm{log}(1-\hat{v}_{k,p}) + \{ \mathrm{log}(\hat{v}_{k,p}) - \mathrm{log}(1-\hat{v}_{k,p}) \} 
    \int dv_{k,p} \ \mathrm{pr}(v_{k,p}|\hat{\mathbf{f}}) v_{k,p} \Bigg] = 0.
\label{eq: C(v_k,vhat_k) final}
\end{equation}
\end{linenomath*}
The solution to Eq.~\eqref{eq: C(v_k,vhat_k) final} is given by 
\begin{linenomath*}
\begin{align}
    \hat{v}_{k,p}^{*}  =  \int d v_{k,p} \ \mathrm{pr}( v_{k,p} | \hat{\mathbf{f}})  v_{k,p},
\label{eq: posterior-mean estimate}
\end{align}
\end{linenomath*}
which is simply the posterior-mean estimate of $v_{k,p}$. 
Thus, an optimization procedure that minimizes Eq.~\eqref{eq: C(v_k,vhat_k) after simp} yields the posterior-mean estimate of the true fractional volume occupied by the left and right caudate, putamen, and GP, within each voxel of the SPECT image. 
Note that the same estimator is obtained when ensemble mean square error is defined as the cost function. 
Thus, the posterior-mean estimator also yields the lowest mean square error among all possible estimators. 
Further, this estimator is unbiased in a Bayesian sense \cite{liu2021bayesian}. 

\subsection{Implementation of the proposed method}
\label{sec: Methods (tissue-fraction estimation technique (implementation))}
Minimizing Eq.~\eqref{eq: C(v_k,vhat_k) after simp} requires sampling from the posterior distribution, $\mathrm{pr}(v_{k,p}|\mathbf{\hat{f}})$. 
However, such sampling is challenging due to the high dimensionality and unknown analytical form of this distribution. 
To address this challenge, the proposed method was implemented using a supervised DL-based approach. 
Specifically, an encoder-decoder-based neural network was constructed. 
During training, the network was input a population of $3$-D SPECT images and the corresponding true fractional volumes occupied by the caudate, putamen, GP, and background, i.e., the $\{\mathbf{v}_k\}$. 
The network was then trained to minimize Eq.~\eqref{eq: C(v_k,vhat_k) after simp} using the Adam algorithm \cite{kingma2014adam} to yield a posterior-mean estimate of $\{\mathbf{v}_k\}$ given an input SPECT image.

\begin{figure*}[ht]
    \centering
    \includegraphics[width=\textwidth]{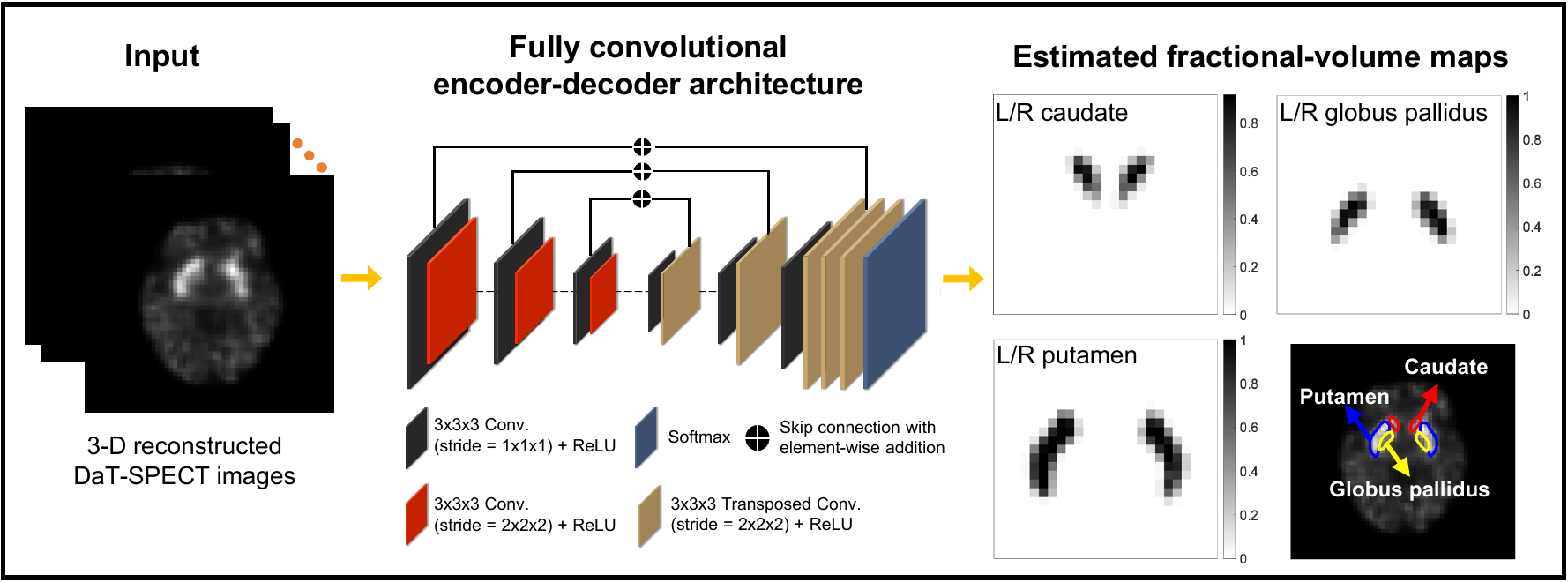}
    \caption{Implementation of the proposed method using a supervised DL-based approach. (L/R: left/right)}
    \label{fig: implementation of tissue-fraction estimation technique}
\end{figure*}

The architecture of the network is similar to that of networks designed for estimation tasks, such as image denoising \cite{creswell2017denoising} and reconstruction \cite{nath2020deep}. 
To summarize, the network comprises an encoder and a decoder. 
The encoder extracts local spatial features from the input SPECT image and the decoder maps the extracted features to the fractional volumes. 
In the final layer, the network outputs the fractional volumes occupied by the left and right caudate, putamen, and GP, and the background, within each voxel of the input SPECT image.
To stabilize the network training, skip connections with element-wise addition were applied between the output of layers in the encoder and decoder. 
Additionally, dropout was applied to prevent overfitting \cite{srivastava2014dropout}. 
Further, instead of performing slice-by-slice or patch-based training \cite{kamnitsas2017efficient}, the network was designed to be input the whole $3$-D image to learn the maximal amount of global contextual information.
Implementation of the proposed method is illustrated in Fig.~\ref{fig: implementation of tissue-fraction estimation technique}, with a detailed description of the network architecture provided in table~\ref{tab: network architecture}.

\begin{table}[]
\centering
\caption{Network architecture. (Conv.: convolutional)}
\label{tab: network architecture}
\resizebox{\textwidth}{!}{%
\begin{tabular}{@{}ccccccc@{}}
\toprule
Layer  & Layer type       & \# filters & Filter size         & Stride size         & Input size                          & Output size                         \\ \midrule
1      & Conv.            & 32         & 3$\times$3$\times$3 & 1$\times$1$\times$1 & 128$\times$128$\times$128$\times$1  & 128$\times$128$\times$128$\times$32 \\
2      & Conv.            & 32         & 3$\times$3$\times$3 & 2$\times$2$\times$2 & 128$\times$128$\times$128$\times$32 & 64$\times$64$\times$64$\times$32    \\
3      & Conv.            & 64         & 3$\times$3$\times$3 & 1$\times$1$\times$1 & 64$\times$64$\times$64$\times$32    & 64$\times$64$\times$64$\times$64    \\
4      & Conv.            & 64         & 3$\times$3$\times$3 & 2$\times$2$\times$2 & 64$\times$64$\times$64$\times$64    & 32$\times$32$\times$32$\times$64    \\
5      & Conv.            & 128        & 3$\times$3$\times$3 & 1$\times$1$\times$1 & 32$\times$32$\times$32$\times$64    & 32$\times$32$\times$32$\times$128   \\
6      & Conv.            & 128        & 3$\times$3$\times$3 & 2$\times$2$\times$2 & 32$\times$32$\times$32$\times$128   & 16$\times$16$\times$16$\times$128   \\
7      & Conv.            & 256        & 3$\times$3$\times$3 & 1$\times$1$\times$1 & 16$\times$16$\times$16$\times$128   & 16$\times$16$\times$16$\times$256   \\
8      & Transposed Conv. & 128        & 3$\times$3$\times$3 & 2$\times$2$\times$2 & 16$\times$16$\times$16$\times$256   & 32$\times$32$\times$32$\times$128   \\
9      & Add Layer 5      & -          & -                   & -                   & 32$\times$32$\times$32$\times$128   & 32$\times$32$\times$32$\times$128   \\
10     & Conv.            & 128        & 3$\times$3$\times$3 & 1$\times$1$\times$1 & 32$\times$32$\times$32$\times$128   & 32$\times$32$\times$32$\times$128   \\
11     & Transposed Conv. & 64         & 3$\times$3$\times$3 & 2$\times$2$\times$2 & 32$\times$32$\times$32$\times$128   & 64$\times$64$\times$64$\times$64    \\
12     & Add Layer 3      & -          & -                   & -                   & 64$\times$64$\times$64$\times$64    & 64$\times$64$\times$64$\times$64    \\
13     & Conv.            & 64         & 3$\times$3$\times$3 & 1$\times$1$\times$1 & 64$\times$64$\times$64$\times$64    & 64$\times$64$\times$64$\times$64    \\
14     & Transposed Conv. & 32         & 3$\times$3$\times$3 & 2$\times$2$\times$2 & 64$\times$64$\times$64$\times$64    & 128$\times$128$\times$128$\times$32 \\
15     & Add Layer 1      & -          & -                   & -                   & 128$\times$128$\times$128$\times$32 & 128$\times$128$\times$128$\times$32 \\
16     & Conv.            & 32         & 3$\times$3$\times$3 & 1$\times$1$\times$1 & 128$\times$128$\times$128$\times$32 & 128$\times$128$\times$128$\times$32 \\
17     & Conv.            & 7          & 3$\times$3$\times$3 & 1$\times$1$\times$1 & 128$\times$128$\times$128$\times$32 & 128$\times$128$\times$128$\times$7  \\
Output & Softmax          & -          & -                   & -                   & 128$\times$128$\times$128$\times$7  & 128$\times$128$\times$128$\times$7  \\ \bottomrule
\end{tabular}%
}
\end{table}

Note that while training the proposed method, the ground truth is defined as the fractional volumes occupied by each region within each voxel of the input SPECT image. 
Thus, the method is specifically designed and trained to model the TFEs while performing segmentation.
In contrast, in conventional DL-based segmentation methods \cite{lin2020deep,leung2020physics}, the ground truth is defined such that each voxel is assigned as belonging to a specific region. 
Thus, these methods are inherently limited in modeling the TFEs. 
Further, as stated in Sec.~\ref{sec: introduction}, while these methods can output a probabilistic estimate of each voxel belonging to a region, this probability is unrelated to the TFEs.

\section{Evaluation}

The proposed method was evaluated based on two criteria.
The first criterion was to assess the spatial overlap and shape similarity between the true and estimated segmentations of the caudate, putamen, and GP.
The second criterion was to evaluate the proposed method on the task of quantifying the DaT uptake within those segmented regions.

The description of this evaluation study is grouped into five subparts, namely (a) data collection, (b) network training, (c) testing procedure, (d) process to extract task-specific information, and (e) figures of merit. This is similar to a recently proposed framework for objective evaluation of AI-based methods for medical imaging \cite{jha2021objective}. 

\subsection{Data collection}
\label{sec:evaluation(data collection)}

Evaluation on both the criteria stated above requires the knowledge of the corresponding ground truth. 
Specifically, evaluation on the first criterion requires access to the true segmentations. 
Typically, manual segmentations are used as a surrogate for ground truth.
However, as described in Sec.~\ref{sec: introduction}, manual segmentations are erroneous due to the PVEs in SPECT and suffer from intra- and inter-reader variabilities.
Further, the GP is visually almost impossible to manually delineate from DaT-SPECT images. 
Similarly, evaluation on the second criterion requires the knowledge of true regional uptake, which are not available in clinical studies.

This issue of the unavailability of ground truth when evaluating quantitative-imaging (QI) methods is often addressed by conducting clinically realistic simulation studies \cite{du2005partial,ouyang2006fast,du2009quantitative,jin2013evaluation,jha2016no}.
These studies provide a rigorous mechanism to evaluate QI methods due to the ability to accurately model imaging-system physics and account for patient-population variabilities. 
Thus, the proposed method was evaluated using clinically guided highly realistic simulation studies.

A total of $580$ T$1$-weighted MR images of cognitively healthy individuals were obtained from the OASIS-3 database \cite{lamontagne2019oasis}. 
From these MR images, the caudate, putamen, and GP in both left and right hemispheres of the brain were segmented using the Freesurfer software \cite{fischl2012freesurfer}. 
Segmentations were performed in Montreal Neurological Institute (MNI) space and then transformed back into the native space of each patient, yielding $580$ anatomical templates. 
The dimension of each template was $256\times256\times256$, with a voxel size of $1~\mathrm{mm} \times 1~\mathrm{mm} \times 1~\mathrm{mm}$. 
From each template, the supports of the regions of caudate, putamen, and GP, i.e., the $\boldsymbol{\phi}_k^{MR}$ given by Eq.~\eqref{eq: phi_kn^MR}, were obtained. 
Clinically realistic DaT-tracer distributions within these regions were next simulated.
For each template, the DaT specific binding ratio (SBR) within each region was independently sampled from a Gaussian distribution with mean and standard deviation values obtained from clinical data \cite{son2016imaging}, as presented in table~\ref{tab:SBR_stats}. 
This sampling process ensured the clinically realistic variability of regional DaT uptake in the simulated patient population. 
By the end of this process, a digital phantom-population of $580$ patients were obtained with anatomical templates derived directly from clinical MR images and DaT-tracer distributions guided by clinical data.

The ground-truth tracer distributions were used to generate realistic DaT-SPECT projection data. 
A GE Discovery $670$ scanner (GE Healthcare, Haifa, Israel) with a low-energy high-resolution collimator was simulated using SIMIND, a well validated Monte-Carlo-based simulation software \cite{ljungberg2012simind,morphis2021validation}.
The simulation modeled all relevant image-degrading processes in SPECT, including photon attenuation and scatter, the finite extent of the collimator with a thickness of $3.5~\mathrm{cm}$, the finite energy resolution of $9.8\%$ at $159$ keV, and the intrinsic spatial resolution of $0.39~\mathrm{cm}$ for the detector. 


\begin{table}[]
\centering
\caption{The mean and standard deviation of the specific binding ratio of the left/right caudate, putamen, and globus pallidus, with the occipital region being the reference region.}
\label{tab:SBR_stats}
\resizebox{0.8\textwidth}{!}{%
\begin{tabular}{@{}cclll@{}}
\toprule
Region of interest          & Mean (standard deviation) specific binding ratio &  &  &  \\ \midrule
Left/right caudate          & 3.283 (0.802)                                    &  &  &  \\
Left/right putamen          & 2.920 (0.458)                                    &  &  &  \\
Left/right globus pallidus  & 1.856 (0.398)                                    &  &  &  \\ \bottomrule
\end{tabular}%
}
\end{table}

The EANM/SNMMI practice guideline \cite{morbelli2020eanm} was followed to simulate the DaT-SPECT acquisition.
The protocol was set up to acquire $120$ projection views that provided $360^{\circ}$ coverage of the head. 
The rotational radius was set to $12.8~\mathrm{cm}$.
A zoom factor was adjusted to achieve a voxel size of $4~\mathrm{mm}$.
The number of total detected counts was scaled to clinically realistic value of $2$ million, to which Poisson noise was added. 
This noisy data was then reconstructed using a $3$-D ordered-subset expectation-maximization (OSEM)-based algorithm with $4$ iterations and $8$ subsets. 
The reconstruction compensated for attenuation, scatter, and collimator-detector response. 
The dimension of the reconstructed image was $128\times128\times128$, with a voxel size of $4~\mathrm{mm} \times 4~\mathrm{mm} \times 4~\mathrm{mm}$.

\begin{figure*}[h!]
    \centering
    \includegraphics[width=\textwidth]{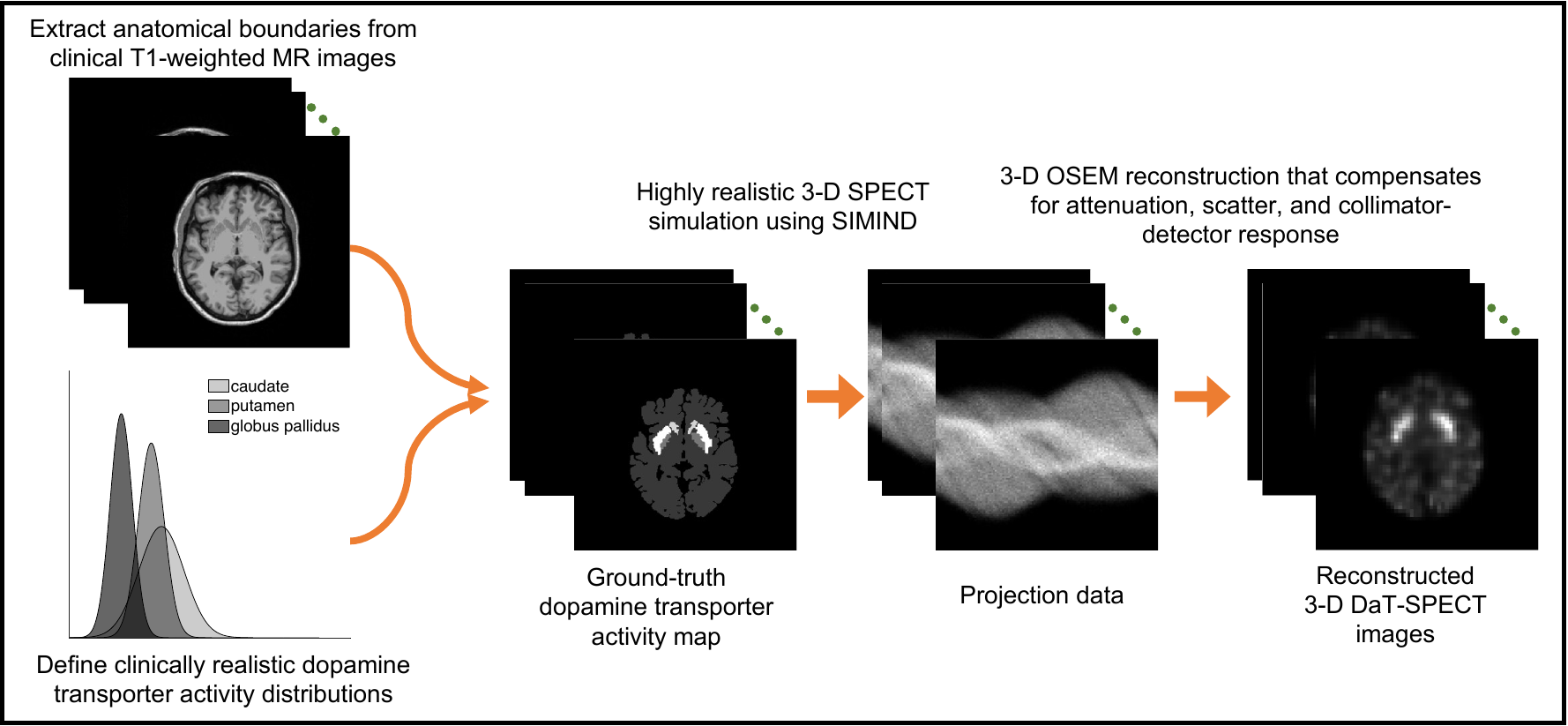}
     \caption{Generation of clinically realistic simulated SPECT projection data and reconstructed images.}    
     \label{fig:simulation}
\end{figure*}

\subsection{Network training}
\label{sec:evaluation(network training)}

Following the simulation procedure (Fig.~\ref{fig:simulation}) described in Sec~\ref{sec:evaluation(data collection)}, a total of 580 reconstructed $3$-D DaT-SPECT images were generated. 
Of these images, $480$ were used for network training.
The hyperparameters of the network were optimized via $5$-fold cross validation.
The remaining $100$ images were reserved for evaluating the performance of the trained network.

As stated in Sec.~\ref{sec: Methods (tissue-fraction estimation technique (implementation))}, the training strategy for the proposed method was designed to model the TFEs while performing segmentation. 
Thus, the ground truth was defined as the fractional volumes occupied by the caudate, putamen, GP, and background within each voxel of a SPECT image.
For each patient, the Freesurfer-defined segmentations from the MR images provided the high-resolution support of each region, i.e., $\boldsymbol{\phi}_{k}^{MR}$ given by Eq.~\eqref{eq: phi_kn^MR}.
From this support, we obtained the true fractional volumes using Eq.~\eqref{eq: v_{k,p} MR}. 
The network was then trained by minimizing the cost function defined in Eq.~\eqref{eq: C(v_k,vhat_k) after simp}. 

\subsection{Testing procedure 1: evaluating the proposed method based on spatial overlap and shape similarity}
\label{sec:evaluation(testing procedure 1)}
In this study, the proposed method was evaluated based on spatial overlap and shape similarity between the true and estimated segmentations. 
Additionally, the proposed method was compared to several commonly used segmentation methods. 
Common methods for segmenting SPECT images can be categorized into those based on thresholding, boundary detection, stochastic modeling, and learning \cite{foster2014review}. 
In this study, $40\%$ SUV-max thresholding \cite{king1991SPECT}, Snakes \cite{kass1988snakes}, Markov random fields-Gaussian mixture model (MRF-GMM) \cite{jha2010clustering}, and a state-of-the-art U-net-based method \cite{leung2020physics} were chosen in each of those categories.
This U-net-based method, similar to common DL-based methods, was designed and trained to classify each image voxel as belonging to a specific region. 

While performing segmentation, the $40\%$ SUV-max thresholding, Snakes, and MRF-GMM methods were provided manual inputs in the form of a seed voxel and/or an initial boundary estimate. 
In contrast, the proposed method and the U-net-based method were fully automated.

\subsection{Testing procedure 2: evaluating the proposed method on the task of quantifying regional uptake}
\label{sec:evaluation(testing procedure 2)}
The goal of segmentation in DaT SPECT is to quantify the uptake within the segmented regions.
Thus, objective evaluation of a segmentation method should assess the efficacy of the method in performing this task \cite{jha2012task}.
For this purpose, we developed a framework to objectively evaluate the proposed method on this quantification task.
An important component of any task-based evaluation is the process to extract the task-specific information \cite{barrett2013foundations,jha2021objective}.
For this evaluation, it is necessary that the procedure to quantify regional uptake itself is optimal when the true boundaries of the considered regions are known, so that the performance on the quantification task directly reflects the accuracy of a segmentation method.

A common procedure to perform such a quantification task is to quantify the regional uptake from the reconstructed images.
However, this procedure is sub-optimal since previous studies have shown that the estimates obtained with clinically used OSEM-based reconstruction methods can be biased \cite{boellaard2001experimental} and/or imprecise \cite{cloquet2012mlem}.
We have also observed from our study that such reconstruction-based procedures led to error in estimated regional uptake, even when the true boundaries were known.  
Thus, this procedure is not suitable to objectively evaluate segmentation methods. 
In this context, one optimal procedure is projection-domain quantification (PDQ).
The PDQ is a maximum-likelihood (ML) technique that operates directly on the projection data to quantify the regional uptake \cite{carson1986maximum}.
By operating directly on the projection data, the PDQ technique accurately models the statistics of noise while performing quantification.
Under the assumption that the uptake within each considered region is homogeneous \cite{carson1986maximum,li2021projection}, the tracer distribution $f(\mathbf{r})$ can be written in terms of $\phi_k(\mathbf{r})$ (Eq.~\ref{eq: phi_k(r)}):
\begin{linenomath*}
\begin{align}
    f(\mathbf{r}) = \sum_{k=1}^K \lambda_k \phi_k(\mathbf{r}).
\label{eq: f(r)}
\end{align}
\end{linenomath*}
When $\phi_k(\mathbf{r})$ is known, the output of the PDQ technique has been observed to be unbiased with a variance that approaches the theoretical lower limit as defined by the Cramér-Rao lower bound \cite{li2021projection}.
Thus, the PDQ technique is optimal for the task of regional uptake quantification. 

When $\phi_k(\mathbf{r})$ is unknown but can be approximated using a segmentation method, any error in the output of the PDQ technique would purely be attributed to the error in approximating $\phi_k(\mathbf{r})$.
Since that output should otherwise be unbiased with minimum variance when $\phi_k(\mathbf{r})$ is known, the PDQ technique provides a mechanism for objective task-based evaluation of the proposed segmentation method. 

We next describe how the PDQ technique was used to quantify the regional uptake from the segmentations yielded by the proposed method. 

\subsubsection{Process to extract task-specific information}
\label{sec:evaluation(testing procedure 2 -- process to extract task-specific information}
As described in Sec.~\ref{sec: Methods (tissue-fraction estimation technique)}, the proposed method estimates the fractional volume that the $k^\mathrm{th}$ region occupies within the $p^\mathrm{th}$ voxel of a SPECT image, i.e., the $\hat{v}_{k,p}$. 
Then, ${\phi_k(\mathbf{r})}$ can be approximated in terms of $\hat{v}_{k,p}$ as follows:
\begin{linenomath*}
\begin{align}
    {\phi_k}_a(\mathbf{r}) =  \sum_{p=1}^P \psi^S_p(\mathbf{r}) \hat{v}_{k,p},
    \label{eq: phi^a_k(r)}
\end{align}
\end{linenomath*}
where ${\phi_k}_a(\mathbf{r})$ denotes the approximation to ${\phi_k}(\mathbf{r})$.
Using Eq.~\eqref{eq: phi^a_k(r)}, $f(\mathbf{r})$ in Eq.~\eqref{eq: f(r)} can be approximated in terms of $\hat{v}_{k,p}$:
\begin{linenomath*}
\begin{align}
\begin{split}
    f_a(\mathbf{r}) &= \sum_{k=1}^K \lambda_k {\phi_k}_a(\mathbf{r}) \\
    &= \sum_{k=1}^K \lambda_k \sum_{p=1}^P \psi^S_p(\mathbf{r}) \hat{v}_{k,p},
\end{split}
\label{eq: f_a(r)}
\end{align}
\end{linenomath*}
where again $f_a(\mathbf{r})$ denotes the approximation to $f(\mathbf{r})$.
Changing the order of summation in Eq.~\eqref{eq: f_a(r)} yields
\begin{linenomath*}
\begin{align}
    f_a(\mathbf{r}) = \sum_{p=1}^P \psi^S_p(\mathbf{r}) \sum_{k=1}^K \lambda_k \hat{v}_{k,p},
\label{eq: f_a(r) change order}
\end{align}
\end{linenomath*}
where the inner summation computes the mean uptake contributed by all the $K$ regions within the $p^\mathrm{th}$ voxel of the reconstructed SPECT image. 
To quantify $\boldsymbol{\lambda}$, the measured projection data $\mathbf{g}$ in Eq.~\eqref{eq: fhat} is modeled as a Poisson-distributed random vector.
Denote the mean vector of $\mathbf{g}$ by $\bar{\mathbf{g}}$, such that
\begin{linenomath*}
\begin{align}
\begin{split}
    \bar{\mathbf{g}} &= \int d^3\mathbf{r} \ h_m(\mathbf{r}) f_a(\mathbf{r}) \\ 
    &= \sum_{k=1}^K \lambda_k \int d^3\mathbf{r} \ h_m(\mathbf{r})  \sum_{p=1}^P \psi^S_p(\mathbf{r}) \hat{v}_{k,p}.
\end{split}
\label{eq: gbar = Hfa}
\end{align}
\end{linenomath*}
Eq.~\eqref{eq: gbar = Hfa} can be written in vector notation as
\begin{linenomath*}
\begin{align}
    \bar{\mathbf{g}} = \mathbf{H} \boldsymbol{\lambda},
    \label{eq: gbar}
\end{align}
\end{linenomath*}
where $\mathbf{H}$ is a system matrix of dimension $M \times K$, with the $(m, k)^\mathrm{th}$ entry given by
\begin{linenomath*}
\begin{align}
    H_{mk} = \int d^3\mathbf{r} \ h_m(\mathbf{r})  \sum_{p=1}^P \psi^S_p(\mathbf{r}) \hat{v}_{k,p}.
\label{eq: PDQ system matrix}
\end{align}
\end{linenomath*}

Denote the probability of measuring $\mathbf{g}$ given $\boldsymbol{\lambda}$ by $\mathrm{pr}(\mathbf{g}|\boldsymbol{\lambda})$. 
Since the counts detected in different projection bins are independent of each other, the likelihood of observing all the $M$ measurements is given by
\begin{linenomath*}
\begin{align}
    \mathrm{pr}(\mathbf{g}|\boldsymbol{\lambda}) = 
    \prod_{m=1}^M \exp \left[ - (\mathbf{H} \boldsymbol\lambda)_m \right] \frac{\left[ (\mathbf{H} \boldsymbol\lambda)_m \right] ^{g_m}}{g_m!}.
    \label{eq: pr(g|lambda)}
\end{align}
\end{linenomath*}
An ML estimate of $\boldsymbol{\lambda}$, denoted by $\hat{\boldsymbol{\lambda}}$, can be obtained by maximizing the logarithm of Eq.~\eqref{eq: pr(g|lambda)}:
\begin{linenomath*}
\begin{align}
    \hat{\boldsymbol{\lambda}} = \underset{\boldsymbol{\lambda}}{\mathrm{arg max}} \log \left[ \mathrm{pr}(\mathbf{g}|\boldsymbol{\lambda}) \right].
\label{eq: lambda hat}
\end{align}
\end{linenomath*}
The estimate, $\hat{\boldsymbol{\lambda}}$, is obtained using a maximum-likelihood expectation-maximization (MLEM) algorithm. 
Specifically, by differentiating the logarithm of Eq.~\eqref{eq: pr(g|lambda)} with respect to $\boldsymbol{\lambda}$ and setting the derivative equal to $0$, the $k^\mathrm{th}$ element of $\hat{\boldsymbol{\lambda}}$ at the $t^\mathrm{th}$ iteration is given by
\begin{linenomath*}
\begin{align}
    \hat{\lambda}_k^{(t)} = \hat{\lambda}_k^{(t-1)} \frac{1}{\sum_{m=1}^M H_{mk}} \sum_{m=1}^M \frac{g_m}{(\mathbf{H}\hat{\boldsymbol{\lambda}}^{(t-1)})_{m}} H_{mk}.
    \label{eq: lambda_hat_k(t)}
\end{align}
\end{linenomath*}

From Eq.~\eqref{eq: lambda_hat_k(t)}, obtaining $\hat{\boldsymbol{\lambda}}$ requires computing the system matrix $\mathbf{H}$ for each of the $100$ test patients.
To obtain the $k^\mathrm{th}$ column of $\mathbf{H}$, based on Eq.~\eqref{eq: PDQ system matrix}, each of the $p^\mathrm{th}$ voxel of the SPECT image was assigned DaT uptake of value equal to the estimated fractional volume $\hat{v}_{k,p}$.
From this activity distribution, an $M$-dimensional noise-free projection data was generated using SIMIND \cite{ljungberg2012simind}, by following the procedure described in Sec.~\ref{sec:evaluation(data collection)}. 
This noise-free projection data, after being normalized by the volume of the $k^\mathrm{th}$ region, resulted in the $k^\mathrm{th}$ column of $\mathbf{H}$. 

\subsubsection{Evaluating the quantification performance using the test-patient dataset}
Using the PDQ technique described in Sec.~\ref{sec:evaluation(testing procedure 2 -- process to extract task-specific information}, the estimated uptake within the caudate, putamen, and GP were obtained based on segmentations yielded by the proposed method. 
The quantification performance was then assessed based on the accuracy and overall reliability of the estimated regional uptake over the $100$ test patients. 

We also assessed the quantification performance of the proposed method for different levels of SBR. 
For this evaluation, we focused on the GP region.
Among the $100$ test patients, the minimum and maximum true SBR values of the GP were $0.94$ and $2.73$, respectively. 
We thus split the test-patient dataset into four different ranges of SBR, namely, $\left[0.94, 1.39\right]$, $\left[1.39, 1.83\right]$, $\left[1.83, 2.28\right]$, and $\left[2.28, 2.73\right]$. 
The quantification performance for each range of SBR values was then assessed. 

Next, we assessed the performance of the proposed method on quantifying the uptake within the GP for different levels of contrast between the GP and the putamen. 
The putamen was chosen since it is in proximity to the GP. 
Among the test patients, the minimum and maximum true GP-to-putamen uptake ratios were $0.40$ and $1.08$, respectively. 
We split the test-patient dataset into four different ranges of uptake ratios, namely, $\left[0.40, 0.57\right]$, $\left[0.57, 0.74\right]$, $\left[0.74, 0.91\right]$, and $\left[0.91, 1.08\right]$.
The quantification performance for each range of uptake ratios was then assessed. 

\subsubsection{Comparing to the U-net-based method}
The PDQ technique requires that the delineations of the caudate, putamen, and GP yielded by a segmentation method do not overlap. 
Results (presented later in Fig.~\ref{fig: contour comparison}) show that of the considered methods, only the proposed method and the U-net-based method consistently yielded non-overlapping segmentations of these regions. Thus, this task-based evaluation focused on the comparison between the proposed method and the U-net-based method.
For each test patient, the system matrix was obtained from the segmentations yielded by the U-net-based method. 
This system matrix was then applied in Eq.~\eqref{eq: lambda_hat_k(t)} to quantify the regional uptake using the PDQ technique.

\subsection{Figures of merit}
\label{sec:evaluation(figures of merit)}
To quantify the spatial overlap between the true and estimated segmentations, we recognize that for each voxel of a SPECT image, the proposed method yields continuous-valued estimates of fractional volumes.
The Dice similarity coefficient (DSC) as proposed in Taha and Hanbury \cite{taha2015metrics} provides a figure of merit (FoM) to evaluate spatial overlap when segmentation methods yield such continuous-valued outputs.
We thus used this FoM in this study. 
A higher value of DSC implies a more accurate segmentation performance.

The shape similarity between the true and estimated segmentations was quantified using the Hausdorff distance (HD) \cite{huttenlocher1993comparing}.
To compute the HD requires obtaining the isosurfaces from segmentations.
For this purpose, we followed a strategy described in Liu et al. \cite{liu2021bayesian}. 
Briefly, a topographic map was first constructed for each of the left and right caudate, putamen, and GP, from the corresponding estimated fractional volumes.
This map illustrates the topography of the fractional volumes by means of isosurfaces. 
Similarly, the ground-truth topographic map of each region was constructed from the corresponding true fractional volumes.
From each of the true and estimated topographic maps, the isosurface representing the set of points at which the value of fractional volume equal to $0.5$ was obtained.
The coordinates of these points were then used to compute HD. 
These points were also plotted to visually assess the performance of the proposed method. 
A lower value of HD implies a more accurate segmentation performance.

For both DSC and HD, the mean values with $95\%$ confidence intervals (CIs) were reported.
Paired sample $t$-tests with $p$-value $<$ $0.01$ were performed to assess whether significant differences exist between the different segmentation methods.

The performance on the task of quantifying regional uptake (Sec.~\ref{sec:evaluation(testing procedure 2)}) was evaluated based on accuracy and overall reliability of the estimated regional uptake over the $100$ test patients, using the FoMs of ensemble normalized bias (NB) and ensemble normalized root mean square error (NRMSE), respectively.
Denote the number of patients by $S$. 
Additionally, denote the true and estimated uptake within region $k$ by $\lambda_{k,s}$ and $\hat{\lambda}_{k,s}$, respectively. 
The ensemble NB and NRMSE of the estimated uptake within the $k^\mathrm{th}$ region are given by
\begin{linenomath*}
\begin{align}
\begin{split}
    \text{ensemble} \ \mathrm{NB}_k &= \frac{1}{S} \sum_{s=1}^S \frac{\hat{\lambda}_{k,s} - \lambda_{k,s}}{\lambda_{k,s}}, \\
    \text{ensemble} \ \mathrm{NRMSE}_k &= \sqrt{\frac{1}{S} \sum_{s=1}^S \left( \frac{\hat{\lambda}_{k,s} - \lambda_{k,s}}{\lambda_{k,s}}  \right)^2}.
\end{split}
\end{align}
\end{linenomath*}

\section{Results}

\subsection{Evaluation based on spatial overlap and shape similarity}
\begin{figure}[h]
    \centering
    \includegraphics[width=\columnwidth]{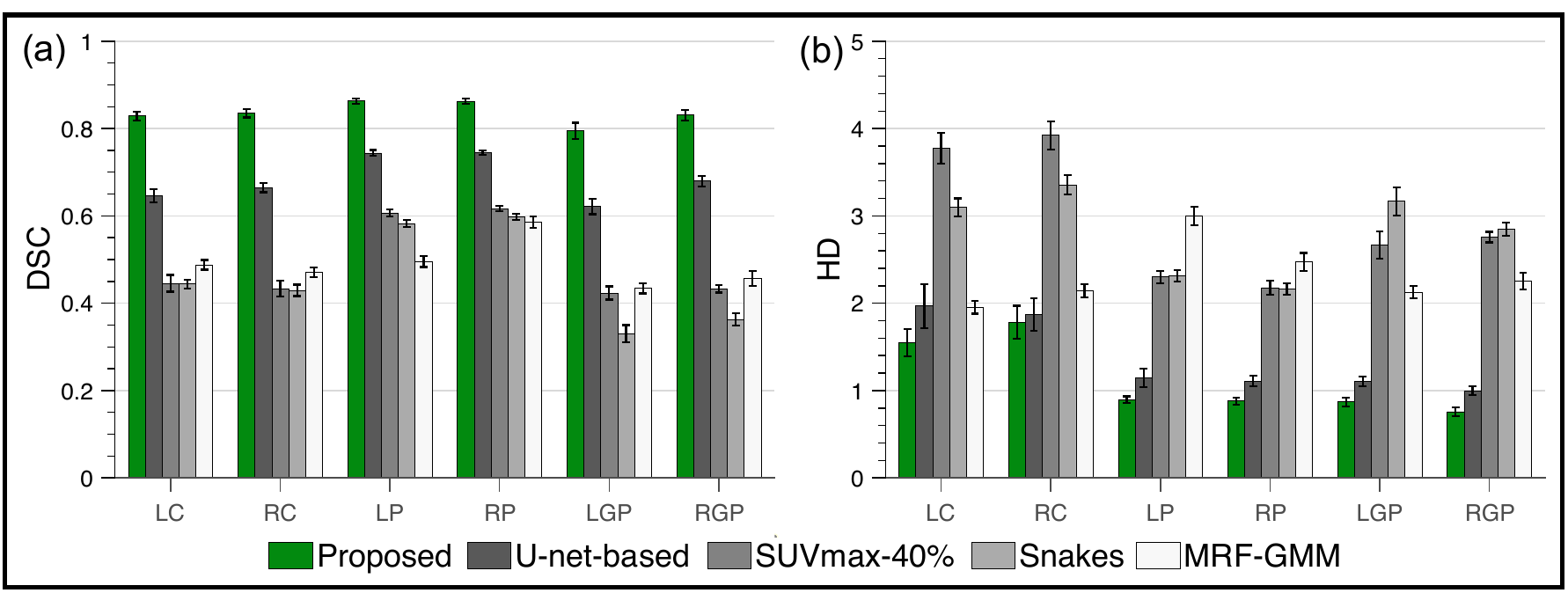}
    \caption{Quantitative comparison of the segmentation performance between the proposed method and other considered methods, on the basis of (a) Dice similarity coefficient (DSC) and (b) Hausdorff distance (HD). LC/RC: left/right caudate; LP/RP: left/right putamen; LGP/RGP: left/right globus pallidus.}
     \label{fig: segm statsitics}
\end{figure}

Quantitatively, the proposed method significantly outperformed ($p < 0.01$) all the other considered SPECT segmentation methods, including the state-of-the-art U-net-based method, based on the criterion of spatial overlap, as quantified using DSC, for all the considered regions (Fig.~\ref{fig: segm statsitics}a).
Further, the proposed method yielded a mean DSC $\sim$ $0.80$ for each considered region, indicating an accurate segmentation performance \cite{zijdenbos1994morphometric}.
Similarly, the proposed method significantly outperformed ($p < 0.01$) all the other considered methods based on the criterion of shape similarity, as quantified using HD, for all the considered regions other than the right caudate, where the performances of the U-net-based method and the proposed method were similar (Fig.~\ref{fig: segm statsitics}b).

\begin{figure}[h!]
    \centering
    \includegraphics[width=\columnwidth]{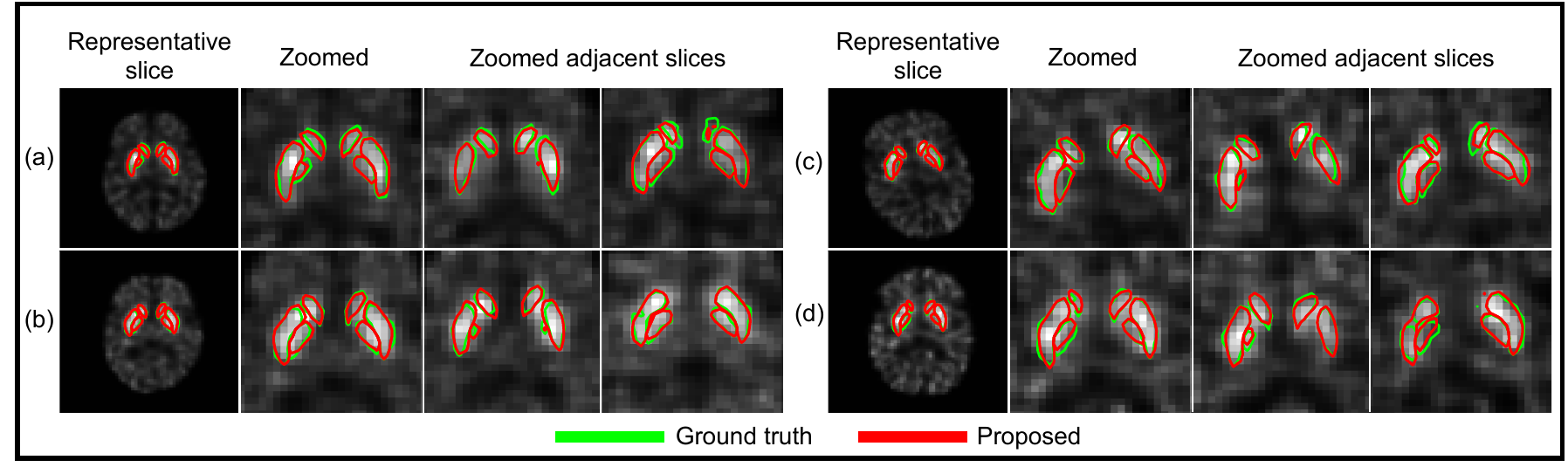}
     \caption{Qualitative evaluation of the proposed method for four representative simulated patients. The SPECT images were displayed at the SPECT resolution.}    
     \label{fig: representative contours}
\end{figure}

\begin{figure}[h!]
    \centering
    \includegraphics[width=\columnwidth]{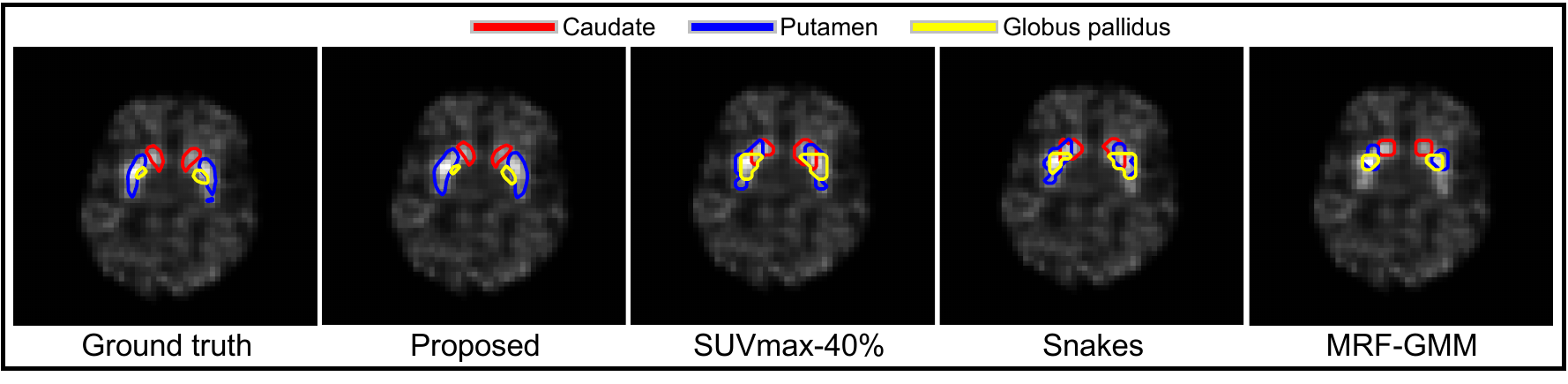}
     \caption{Qualitative comparison between the ground-truth segmentations of caudate, putamen, and globus pallidus and segmentations yielded by the proposed method and conventional computer-aided segmentation methods.}
     \label{fig: contour comparison}
\end{figure}

Representative segmentations yielded by the proposed method, as obtained by following the procedure described in Sec.~\ref{sec:evaluation(figures of merit)}, are shown in Fig.~\ref{fig: representative contours}.
The proposed method was observed to yield accurate segmentation of all considered regions.
Additionally, in Fig.~\ref{fig: contour comparison}, the $40\%$ SUV-max thresholding, Snakes, and MRF-GMM methods yielded highly overlapped segmentations of the caudate, putamen, and GP. 
In contrast, the proposed method provided accurate non-overlapping segmentations of these regions.

\begin{figure}[h!]
    \centering
    \includegraphics[width=\columnwidth]{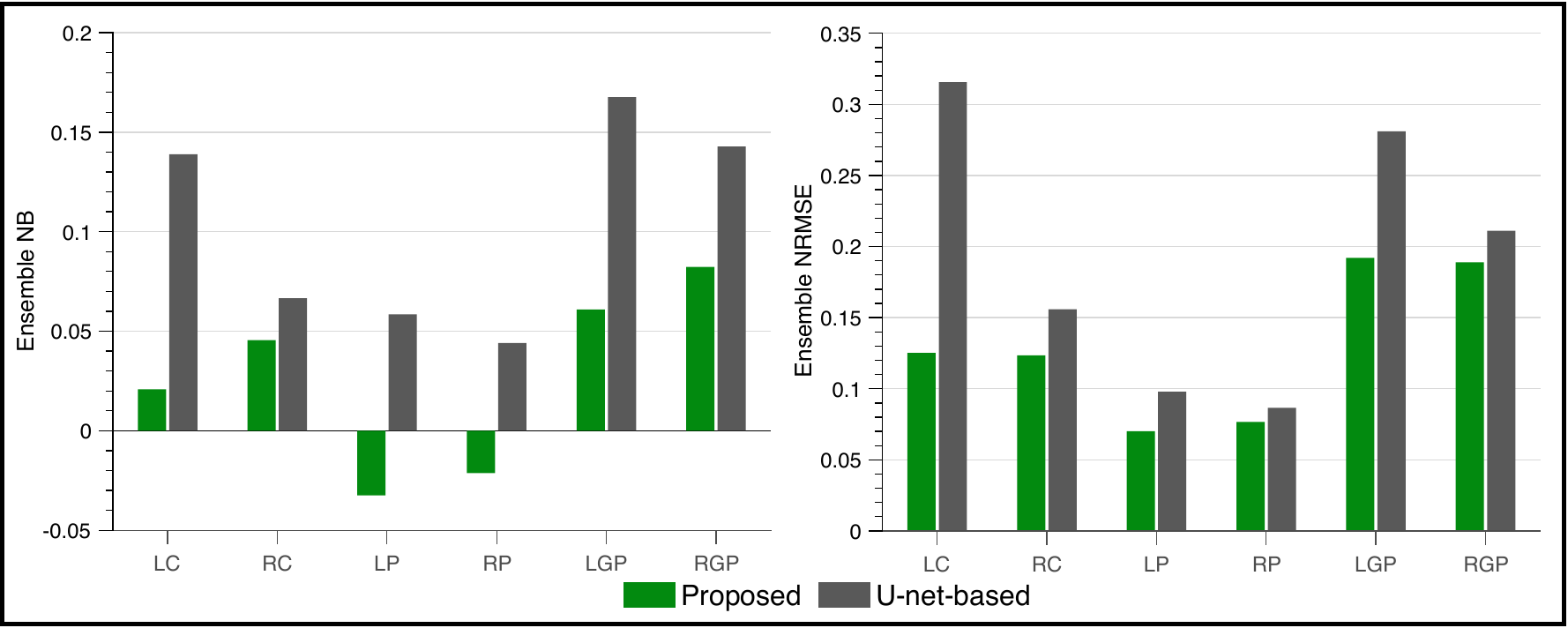}
     \caption{The ensemble normalized bias (NB) and ensemble normalized root mean square error (NRMSE) of the estimated regional uptake obtained with the proposed and U-net-based methods over the $100$ test patients. (LC/RC: left/right caudate; LP/RP: left/right putamen; LGP/RGP: left/right globus pallidus)}
     \label{fig: pdq}
\end{figure}

\begin{figure}[h!]
    \centering
    \includegraphics[width=\columnwidth]{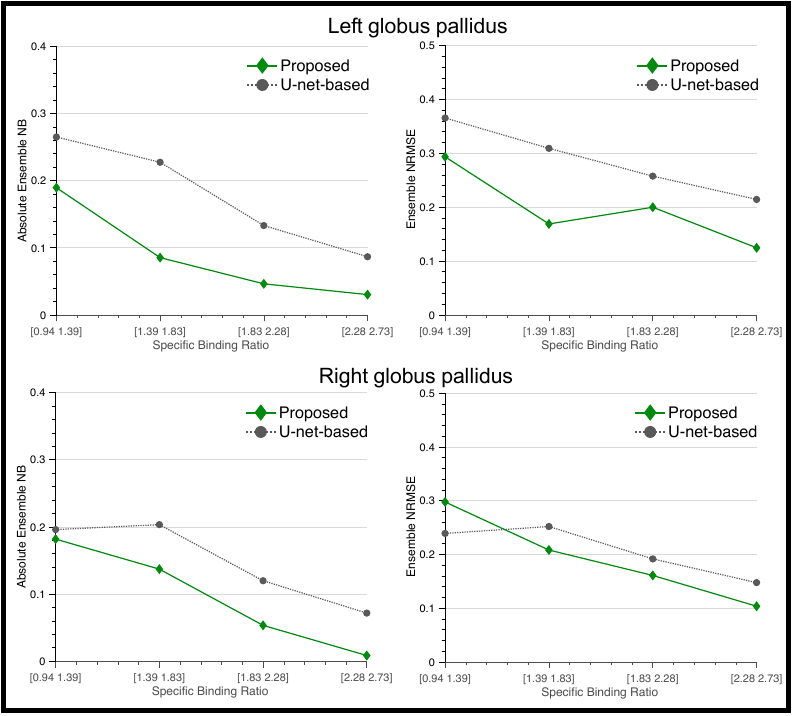}
     \caption{The absolute ensemble normalized bias (NB) and ensemble normalized root mean square error (NRMSE) of the quantified uptake within left/right globus pallidus (GP) using the proposed and U-net-based methods for different ranges of specific binding ratio of the GP.}
     \label{fig: GP_SBR}
\end{figure}

\begin{figure}[h!]
    \centering
    \includegraphics[width=\columnwidth]{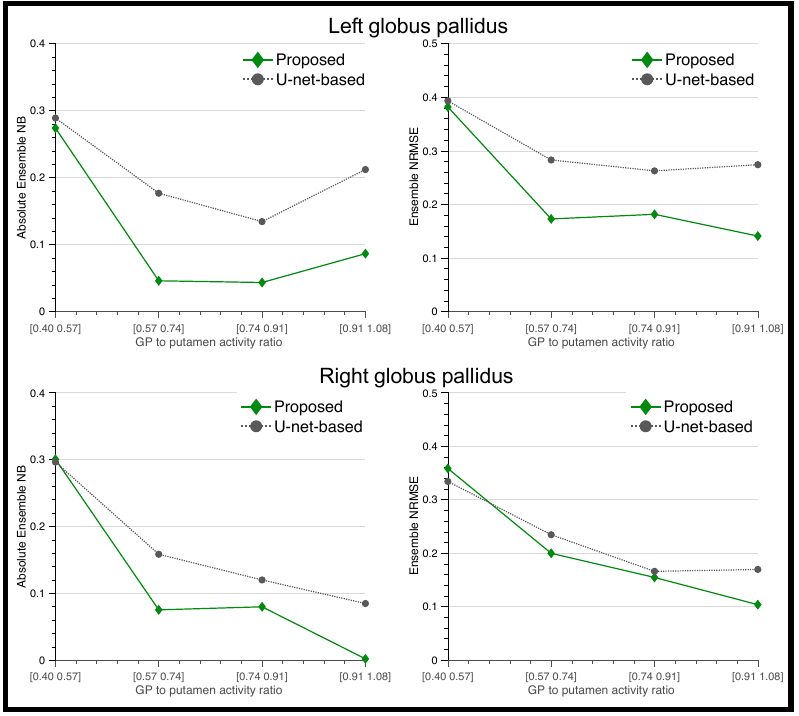}
     \caption{The absolute ensemble normalized bias (NB) and ensemble normalized root mean square error (NRMSE) of the quantified uptake within left/right globus pallidus (GP) using the proposed and U-net-based methods for different GP-to-putamen uptake ratios.}
     \label{fig: GP_activity_ratio}
\end{figure}

\subsection{Evaluation on the task of quantifying regional uptake}

The ensemble NB and NRMSE of the estimated regional uptake yielded by the proposed method are shown in Fig.~\ref{fig: pdq}.
The method reliably quantified the regional uptake with low ensemble NRMSE values ($< 20\%$) for all considered regions. 
In particular, the method yielded even lower ensemble NRMSE values ($\sim 10\%$) for the caudate and putamen.
Further, the method yielded a low ensemble NB value ($< 10\%$) for each region. 
Additionally, the proposed method consistently outperformed the U-net-based method on the basis of both ensemble NB and ensemble NRMSE. 

Fig.~\ref{fig: GP_SBR} shows that for all the considered levels of SBR of the GP, the proposed method reliably quantified the uptake in the GP.
On average, the proposed method yielded an ensemble NRMSE $\sim 20\%$ and absolute ensemble NB $\sim 10\%$. 
We also observe that as the SBR increased, the proposed method yielded improved quantification performance with ensemble NRMSE and absolute ensemble NB approaching $10\%$ and $0\%$, respectively. 
Additionally, the proposed method consistently yielded lower absolute ensemble NB compared to the U-net-based method.

Next, we observe from Fig.~\ref{fig: GP_activity_ratio} that for all the considered GP-to-putamen uptake ratios, the proposed method reliably quantified the uptake within the GP and consistently outperformed the U-net-based method in terms of accuracy, as measured by absolute ensemble NB.
Further, the proposed method yielded lower or similar ensemble NRMSE compared to the U-net-based method. 

\section{Discussion}

Reliable quantification of uptake within the caudate, putamen, and GP requires that these regions are segmented accurately from DaT-SPECT images.
Conventional SPECT segmentation methods often yield limited performance due to their inability to account for PVEs, and in particular, TFEs.
Most of these methods are designed to perform segmentation via voxel-wise classification, which inherently limits their ability to model the TFEs.
To address this limitation, a tissue-fraction estimation-based segmentation method was proposed that estimates the fractional volume occupied by the caudate, putamen, and GP within each voxel of a SPECT image.

Evaluations using clinically guided highly realistic simulation studies show that the proposed method yielded accurate performance on segmenting the caudate, putamen, and GP.
Quantitatively, the method yielded a high mean DSC of $\sim 0.80$ for each region and significantly outperformed ($p < 0.01$) all other considered segmentation methods, including a state-of-the-art U-net-based method (Fig.~\ref{fig: segm statsitics}).
The proposed method was implemented using an encoder-decoder-based neural network architecture similar to that used in the U-net-based method. 
Given the similarity in the network architectures, the superior performance of the proposed method can be attributed to the ability to model the TFEs using a cost function that penalizes the deviation between the true and estimated fractional volumes. 

On the task of quantifying regional uptake, the proposed method reliably quantified the uptake within all considered regions, and outperformed the U-net-based method (Fig.~\ref{fig: pdq}).
Additionally, the proposed method reliably quantified the uptake within the GP for different levels of SBR of the GP and different GP-to-putamen uptake ratios, and frequently outperformed the U-net-based method (Figs.~\ref{fig: GP_SBR}, \ref{fig: GP_activity_ratio}). 
These results emphasize the importance of modeling the TFEs while performing quantification.

A distinctive capability of the proposed method is to accurately segment the small-sized GP and reliably quantify the uptake within this region.
As shown in table~\ref{tab:SBR_stats}, the uptake within GP is typically low. 
When the uptake is closer to that within background region, segmentation can be more challenging. 
We observe that even in such a scenario, the proposed method accurately segmented the GP.
Consequently, the method yielded reliable regional uptake quantification (Fig.~\ref{fig: GP_SBR}).
Additionally, to the best of our knowledge, no validated tools are available to segment the GP from DaT-SPECT images.
The ability to segment the GP is very important as this presents an opportunity to rigorously evaluate pallidal uptake as a biomarker for measuring the severity of PD.
This also opens up a new and important research frontier on evaluating the functional characteristics of the GP in patients with PD.
Additionally, the ability of the proposed method to accurately segment the caudate and putamen provides a mechanism to evaluate quantitative features, including the shape and texture, extracted from these regions and determine whether these features exhibit differences in different forms of parkinsonism. 
All these studies may lead to new biomarkers for measuring the severity of PD and possibly for differential diagnosis of PD.

An important advantage of the proposed method is that the application of this method does not require the image of the same patient from any other modalities.
Another class of methods to segment striatal regions from DaT-SPECT images uses MR images of the same patient \cite{rahmim2016application}. 
However, often times, images from both these modalities may not be available. 
One mechanism to obtain such images may be from SPECT/MR systems \cite{hutton2018development}, but these systems are currently not available in clinic. 
The proposed method, by virtue of having no such requirement, can segment the DaT-SPECT image of a patient even when the MR image is not available. 
This advantage may facilitate the wider adoption of the method. 
However, the method does require the SPECT and MR images of the same patient during the training phase. 
These images could be obtained from databases such as Parkinson's Progression Marker Initiative (PPMI) \cite{marek2011parkinson}. 
Another option is to use simulation-based approaches for pre-training the network, followed by transfer learning with a small number of co-registered SPECT and MR images. 
A similar approach was suggested in Leung et al. \cite{leung2020physics} and was observed to substantially reduce the number of training images.
Investigating the performance of such a transfer-learning-based approach to training is an important area of future research. 

Another contribution of this study is to develop a framework for objective evaluation of segmentation methods on the task of regional uptake quantification. 
Previously, approaches have been suggested to evaluate segmentation methods on quantification tasks \cite{jha2012task,jha2010evaluating}.
Since segmentation is typically an intermediate step for a certain quantification task, such an evaluation can directly quantify the performance of segmentation methods on this task. 
In the developed framework, we used a PDQ technique to conduct this task-based evaluation. 
As outlined in Sec.~\ref{sec:evaluation(testing procedure 2)}, when provided the true boundaries of the segmented regions, the PDQ technique yields an unbiased estimate of the regional uptake with a variance approaching the Cramér-Rao lower bound.
The choice of such an optimal estimator allows us to purely study the impact of segmentation on the quantification performance \cite{jha2021objective}, since any error in quantifying the regional uptake would be attributed to the error in segmentation.

There are some limitations in this study.
First, the proposed method was evaluated using simulation studies. 
While the simulations were designed to be highly realistic and clinically relevant, they may not have modeled all aspects of patient physiology and system instrumentation.
Evaluation using physical-phantom and patient studies can help address these limitations. 
Validation with patient studies requires a gold standard that is not available in DaT-SPECT studies. 
To address this challenge, no-gold-standard evaluation techniques have been developed \cite{kupinski2002estimation, jha2012task, jha2016no, jha2017practical, liu2022no}. 
These techniques would provide a mechanism to evaluate the proposed method with patient data.
A second limitation is that the proposed method focused on segmenting the caudate, putamen, and GP from DaT-SPECT images and considered rest of the brain as background.
However, post-mortem and PET-based studies have identified DaT presence and uptake in other regions of the brain such as nucleus accumbens and substantia nigra \cite{sun2012dopamine,brown2013validation}. 
Further, the GP can itself be separated into two parts, namely the internal and external GP. 
It remains to be determined how these two parts may contribute independently to the progression of PD. 
Results from this study motivate extending the method to segment those regions. 
Finally, in the simulations, we assumed that the uptake within the various regions is homogeneous.
However, the uptake within putamen may be heterogeneous in patients with PD. 
Modeling this heterogeneity and evaluating the ability of the proposed method to segment regions with heterogeneous uptake are important future research directions.
A related limitation is that in the objective task-based evaluation study, the PDQ technique assumes that the uptake within different regions is homogeneous.
Developing methodologies to reliably quantify uptake even when the tracer distribution within the region of interest is heterogeneous can help address this limitation.
While there are ongoing efforts in this area \cite{lim2018pet}, this is another important future research direction.

\section{Conclusion}
A tissue-fraction estimation-based segmentation method was proposed for quantitative DaT SPECT.
The proposed method yields fully-automated segmentation of the caudate, putamen, and globus pallidus by estimating the fractional volumes occupied by these regions within each voxel of a $3$-D DaT-SPECT image.
Evaluations using clinically guided highly realistic simulation studies demonstrated the ability of the method to accurately segment these regions.
Additionally, the proposed method significantly outperformed all other considered segmentation methods, including a state-of-the-art U-net-based method.
Further, on the task of regional uptake quantification, the proposed method reliably estimated the uptake within the considered regions.
Overall, these results demonstrate the efficacy of the proposed method for accurate segmentation and reliable quantification from DaT-SPECT images.
The results motivate further evaluation of the method with physical-phantom and patient studies. 

\section*{Acknowledgements}
Financial support for this work was supported by the National Institute of Biomedical Imaging and Bioengineering R01-EB031051, R01-NS124789, R56-EB028287, and R21-EB024647 (Trailblazer Award), the Dystonia Medical Research Foundation (DMRF), the American Parkinson Disease Association (APDA), the Greater St. Louis Chapter of the APDA, the Barnes-Jewish Hospital Foundation (Elliot Stein Family Fund), and the Paula \& Rodger Riney Fund.

\bibliographystyle{./medphy.bst}
\bibliography{./ref}

\end{document}